\documentstyle[epsfig,citesort]{aipproc}
\def\s{\sigma}
\def\a{\alpha}
\def\g{\gamma}

\newcommand{\re}{\mathop{\mathrm{Re}}}
\def\epem{e^+e^-}
\def\rts{\sqrt s}
\newcommand{\GG}{\mbox{$\gamma\gamma$}}
\def\be{\begin{equation}}
\def\ee{\end{equation}}
\def\ie{{\it i.e.}}

\begin{document}
\title{Pair production of $W$ bosons \\ at the photon linear collider: \\
a window to the electroweak symmetry breaking?}

\author{Georgy~Jikia\footnote{Alexander von Humboldt Fellow;
        e-mail: jikia@phyv4.physik.uni-freiburg.de}}
\address{Albert--Ludwigs--Universit\"{a}t Freiburg, Fakult\"{a}t f\"{u}r Physik\\
Hermann--Herder Str.3, D-79104 Freiburg, Germany \\
and \\
Institute for High Energy Physics, Protvino\\
Moscow Region 142284, Russian Federation}

\maketitle

\begin{abstract}
  Recent progress in calculating ${\cal O}(\a)$ electroweak
  corrections to $W^+W^-$ pair production in photon-photon collisions
  is reviewed. The potential of the Photon Linear Collider to study
  anomalous $W$ couplings is discussed.
\end{abstract}

\section*{Introduction}

Linear colliders offer unique opportunities to study photon-photon
collisions obtained using the process of Compton backscattering of
laser light off electron beams from the linear collider
\cite{Telnov97}. This option is included now in conceptual design
reports of the NLC, JLC and TESLA/SBLC projects \cite{NLC,CDR}.  The
expected physics at the Photon Linear Collider (PLC) is very rich and
complementary to that in $e^+e^-$ collisions (see {\it e.g.}
\cite{Jikia97} and references therein).

Without the discovery of a light Higgs boson at LEP2, LHC or the
linear collider, the best strategy to probe the symmetry breaking
sector would lie in the study of the self couplings of the $W$, $Z$
bosons.  The large cross sections for the processes involving $W$'s
make $\g\g$ colliders especially attractive tool in probing the $W$ self
couplings.  Indeed, the reaction $\g\g\to W^+W^-$ would be the
dominant source of the $W^+W^-$ pairs at future linear colliders,
provided that photon-photon collider option will be realized. The Born
cross section of $W^+W^-$ pair production in photon-photon collisions
in the scattering angle interval $10^\circ < \theta^\pm < 170^\circ$
is 61~pb at $\sqrt{s_{\g\g}}=500$~GeV and 37~pb at 1~TeV.
Corresponding cross sections of $W^+W^-$ pair production in $e^+e^-$
collisions are an order of magnitude smaller: 6.6~pb at 500~GeV and
2.5~pb at 1~TeV.  With more than a million $WW$ pairs per year a
photon-photon collider can be really considered as a $W$-factory and
an ideal place to conduct precision tests on the anomalous triple
\cite{Ano,BBB} and quartic \cite{BeBu,BBB,BB}
couplings of the $W$ bosons.  In addition, in the process of triple
$WWZ$ vector boson production it is possible to probe the tri-linear
$ZWW$ and quartic couplings \cite{BB,eboli,BBB,gamma-gamma2} as well
as the ${\cal C}$ violating anomalous $ZWW$, $\g ZWW$ interactions
\cite{gamma-gamma2}.

With the natural order of magnitude on anomalous couplings one needs
to know the ${\cal SM}$ cross sections with a precision better than 1\% to
extract these small numbers. From a theoretical point of view this
calls for the very careful analysis of at least $\cal{O}(\alpha)$
corrections to the cross section of $W^+W^-$ pair production in $\g\g$
collisions, which were recently calculated including virtual
corrections \cite{DDS-WW} and complete $\cal{O}(\alpha)$ corrections
taking into account both virtual one-loop corrections and real photon
and $Z$-boson emission \cite{Jikia-WW}.

\section*{$\g\g\to W^+W^-$ cross sections and ${\cal O}(\a)$ corrections}

Inclusive cross section of $W^+W^-$ pair production in photon-photon
collisions to third order in $\a$ is given by the IR-finite sum of
Born cross section, interference term between the Born and one-loop
amplitudes and cross section of $W^+W^-$ pair production accompanied
by the real photon emission. At energies above the $WWZ$ threshold one
should add the cross section of $W^+W^-Z$ production.

\begin{eqnarray}
&&d\s(\g\g\to W^+W^-+X)\,=\, 
\nonumber\\
&&d\s^{Born}(\g\g\to W^+W^-)\,+\,
\frac{1}{2s_{\g\g}} 2 \re\Biggl({\cal M}^{Born}
{{\cal M}^{1-loop}}^*\Biggr) dPS^{(2)}
\label{cs} \\
&&+\,d\s^{soft}(\g\g\to W^+W^-\g)\biggr|_{\omega_\g<k_c}
\,+\,d\s^{hard}(\g\g\to W^+W^-\g)\biggr|_{\omega_\g>k_c}\nonumber\\
&&+\,d\s^Z(\g\g\to W^+W^-Z).
\nonumber
\end{eqnarray}

In contrast to the corrections to the $Z$ boson production at LEP1, as
a consequence of the universality of the electric charge no power or
logarithmically enhanced corrections involving $m_H^2/M_W^2$,
$\log(m_H^2)$, $m_t^2/M_W^2$, $\log(m_t^2)$ or $\log(s/m_f^2)$ appear
in the on-shell scheme in the limit, when Higgs-boson or top-quark
masses are much larger than the collision energy or the fermion mass
$m_f$ is much smaller than $\sqrt{s}$ \cite{DDS-WW,Jikia-WW}.

Figure~1 shows total cross section of $WW$ pair production summed over
$WW$ and $WW\g$ final states and integrated over $W^\pm$ scattering
angles in the interval $10^\circ<\theta^\pm<170^\circ$ as a function
of energy for various polarizations.  The bulk of the cross section
originates from transverse $W_TW_T$ pair production.  Transverse $W$'s
are produced predominantly in the forward/backward direction and the
helicity conserving amplitudes are dominating. Cross sections
integrated over the whole phase space are non-decreasing with energy.
For a finite angular cutoff they do decrease as $1/s$, but still they
are much larger than suppressed cross sections. For the dominating
$++++$, $+-+-$, $+--+$ helicity configurations corrections are
negative and they rise with energy ranging from $-3\%$ at 500~GeV to
$-25\%$ at 2~TeV. The correction for the next to the largest cross
section $\s_{+-00}$ is also negative. For the other helicities
radiative corrections are positive at high energies due to dominating
positive contribution of real photon emission.  Radiative corrections
to cross sections $\s_{+-TL}$, which are decreasing at Born level as
$1/s^2$, are positive and large. For the cross sections $\s_{++00}$
and $\s_{+-++}$, which are decreasing as $1/s^3$ for a finite angular
cutoff, corrections are even larger. The cross section $\s_{++--}$ is
decreasing at tree level as $1/s^5$ for a finite angular cutoff and is
quite negligible at high energy.  The cross sections $\s_{++LT}$ and
$\s_{+++-}$ vanish at the Born level, so only the process of $WW\g$
production contributes in Figure~1.  The cross sections $\s_{++--}$
and $\s_{++00}$ exhibit a clear clear Higgs resonance peak and the
interference pattern, respectively, at $\sqrt{s_{\g\g}}=m_H=300$~GeV.

\begin{figure}
\setlength{\unitlength}{1in}
\begin{picture}(5.4,3)
\put(-.1,0){\epsfig{file=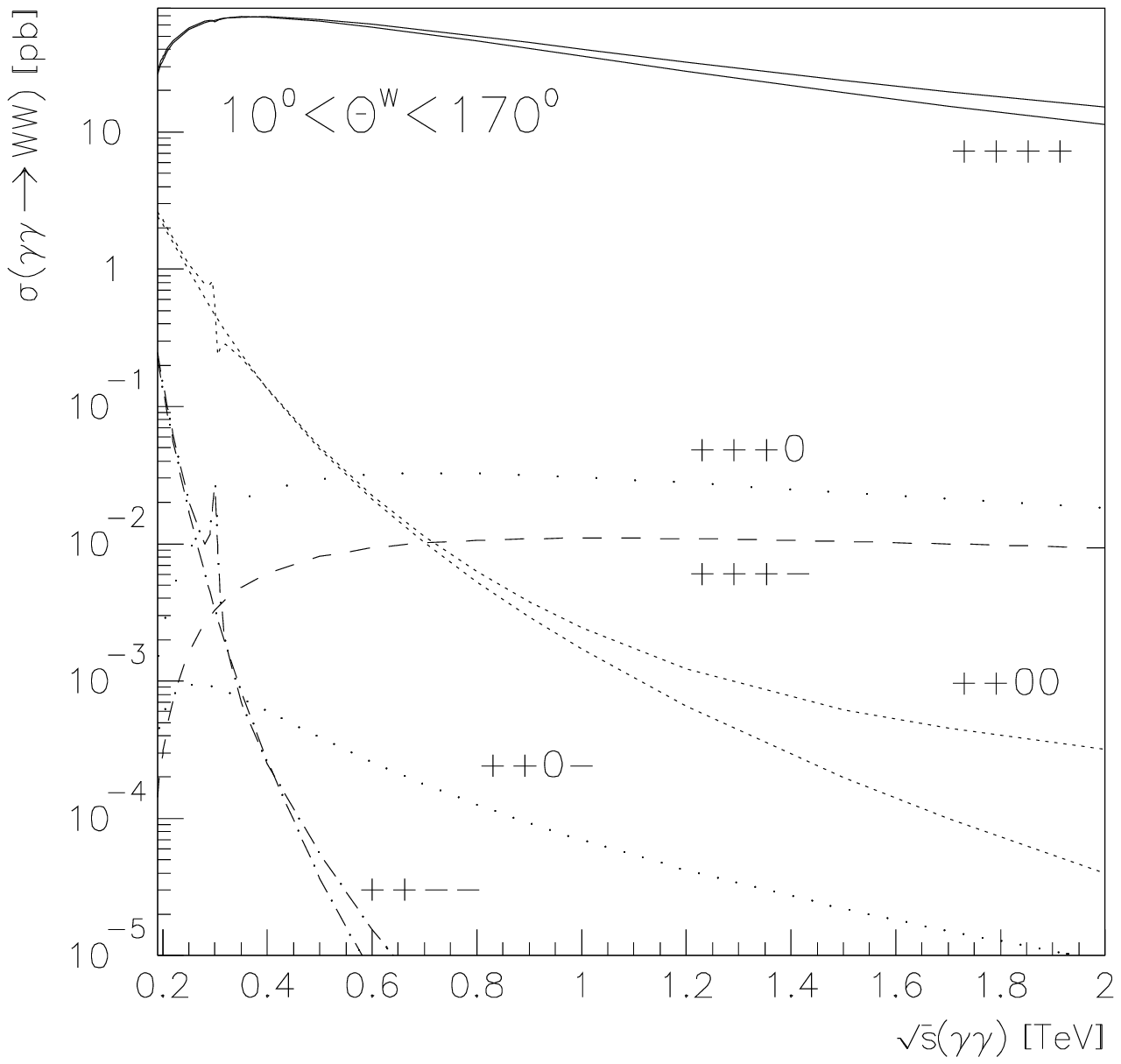,width=3in,height=3in}}
\put(2.7,0){\epsfig{file=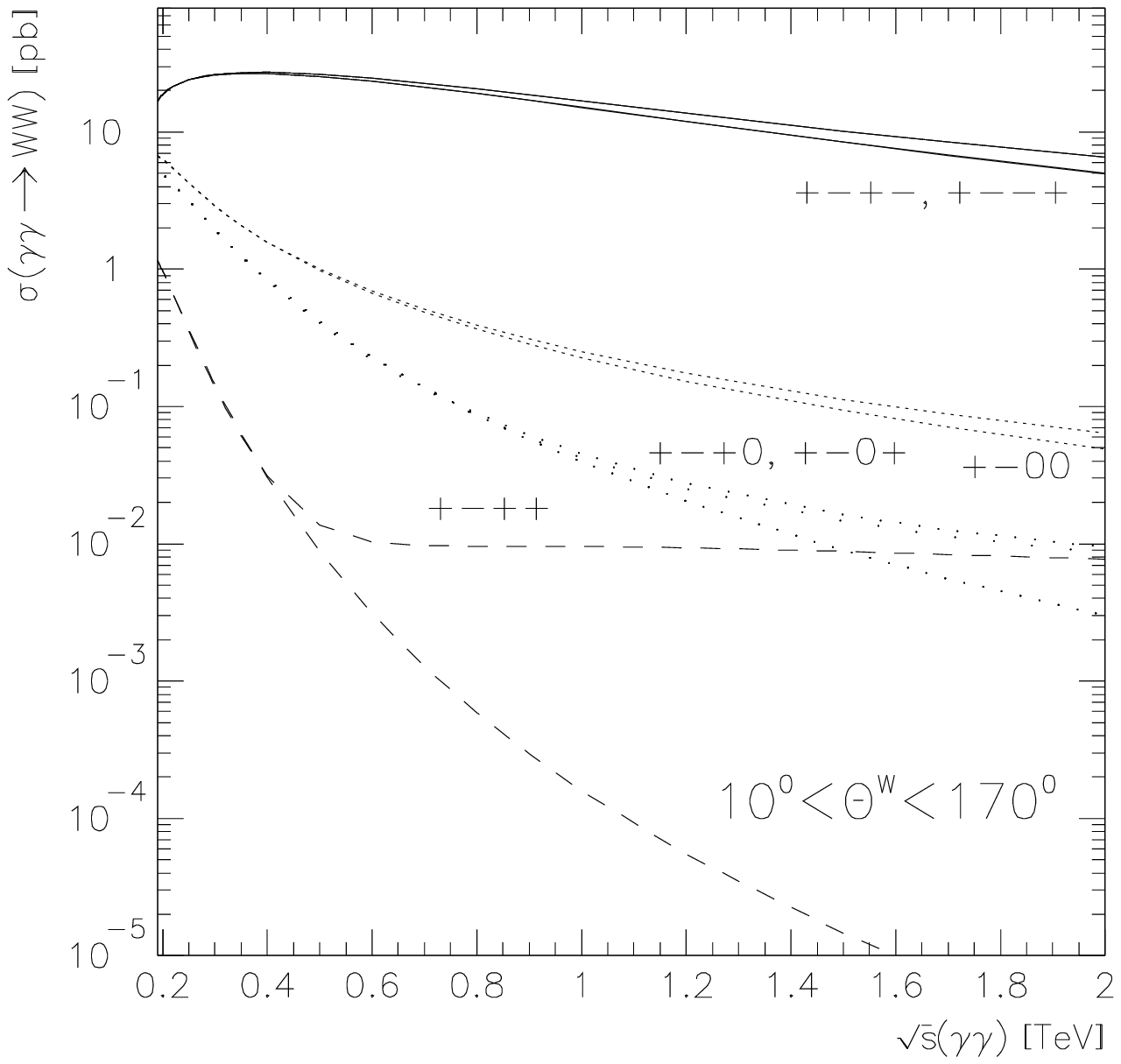,width=3in,height=3in}}
\end{picture}
\caption{Total cross sections of $WW(\g)$ production for various
polarizations.  Born and corrected cross sections are shown. The
curves nearest to the helicity notations represent the corrected cross
sections.}
\end{figure}

\begin{table}[htb]
\caption{Unpolarized Born cross sections and relative
corrections for various intervals of $W^\pm$ scattering
angles. Corrections originating from real hard photon
($\omega_\g>k_c=0.1$~GeV) and $Z$-boson emission as well as 
sum of soft photon and virtual boson contributions, fermion virtual
corrections and total corrections are given separately.}

\begin{tabular}{ccccccc}
&\multicolumn{5}{c}{$\sqrt{s} = 300$~GeV}&\\ \hline
$\theta_{W^\pm}$, ${}^\circ$ & $\sigma^{Born}$, $pb$ & $\delta^{hard}$, \% &
$\delta^{Z}$, \% & $\delta^{soft+bose}$, \% & $\delta^{fermi}$, \% 
&$\delta^{tot}$, \% \\ \hline
$  0^\circ < \theta < 180^\circ$ &  70.22     &  4.15    
& 2.64$\cdot 10^{-2}$& $-$7.09    & 0.327    & $-$1.37    
 \\
$ 10^\circ < \theta < 170^\circ$ &  64.46     &  4.11    
& 2.74$\cdot 10^{-2}$& $-$7.31    & 0.257    & $-$1.59    
 \\
$ 30^\circ < \theta < 150^\circ$ &  38.15     &  4.09    
& 3.27$\cdot 10^{-2}$& $-$8.62    & $-$0.123    & $-$2.67    
 \\
$ 60^\circ < \theta < 120^\circ$ &  12.96     &  4.02    
& 2.94$\cdot 10^{-2}$& $-$10.7    & $-$0.415    & $-$3.75    
\\ \hline
&\multicolumn{5}{c}{$\sqrt{s} = 500$~GeV}&\\ \hline
$\theta_{W^\pm}$, ${}^\circ$ & $\sigma^{Born}$, $pb$ & $\delta^{hard}$, \% &
$\delta^{Z}$, \% & $\delta^{soft+bose}$, \% & $\delta^{fermi}$, \% 
&$\delta^{tot}$, \% \\ \hline
$  0^\circ < \theta < 180^\circ$ &  77.50     &  7.96    & 0.468    
& $-$10.1    & 9.04$\cdot 10^{-2}$& $-$1.63    
 \\
$ 10^\circ < \theta < 170^\circ$ &  60.71     &  7.89    & 0.541    
& $-$10.7    & $-$0.242    & $-$2.52    
 \\
$ 30^\circ < \theta < 150^\circ$ &  21.85     &  8.05    & 0.817    
& $-$13.0    & $-$1.34    & $-$5.50    
 \\
$ 60^\circ < \theta < 120^\circ$ &  5.681     &  8.02    & 0.789    
& $-$14.8    & $-$2.13    & $-$8.12
\\ \hline
&\multicolumn{5}{c}{$\sqrt{s} = 1000$~GeV}&\\ \hline
$\theta_{W^\pm}$, ${}^\circ$ & $\sigma^{Born}$, $pb$ & $\delta^{hard}$, \% &
$\delta^{Z}$, \% & $\delta^{soft+bose}$, \% & $\delta^{fermi}$, \% 
&$\delta^{tot}$, \% \\ \hline
$  0^\circ < \theta < 180^\circ$ &  79.99     &  13.3    &  1.55    
& $-$18.7    & $-$5.51$\cdot 10^{-2}$& $-$3.89    
 \\
$ 10^\circ < \theta < 170^\circ$ &  37.04     &  13.4    &  2.39    
& $-$22.6    & $-$1.28    & $-$8.10    
 \\
$ 30^\circ < \theta < 150^\circ$ &  6.924     &  14.2    &  3.96    
& $-$32.1    & $-$3.80    & $-$17.8    
 \\
$ 60^\circ < \theta < 120^\circ$ &  1.542     &  14.2    &  3.88    
& $-$37.1    & $-$5.13    & $-$24.1    
\\   \hline
&\multicolumn{5}{c}{$\sqrt{s} = 2000$~GeV}&\\ \hline
$\theta_{W^\pm}$, ${}^\circ$ & $\sigma^{Born}$, $pb$ & $\delta^{hard}$, \% &
$\delta^{Z}$, \% & $\delta^{soft+bose}$, \% & $\delta^{fermi}$, \% 
&$\delta^{tot}$, \% \\ \hline
$  0^\circ < \theta < 180^\circ$ &  80.53     &  19.0    &  2.91    
& $-$27.2    & $-$7.45$\cdot 10^{-2}$& $-$5.33    
 \\
$ 10^\circ < \theta < 170^\circ$ &  14.14     &  20.1    &  6.38    
& $-$41.6    & $-$2.99    & $-$18.1    
 \\
$ 30^\circ < \theta < 150^\circ$ &  1.848     &  21.5    &  9.77    
& $-$60.1    & $-$6.54    & $-$35.4    
 \\
$ 60^\circ < \theta < 120^\circ$ & 0.3936     &  21.6    &  9.60    
& $-$67.6    & $-$8.04    & $-$44.5    
\end{tabular}
\end{table}

In Table~1 Born cross sections and relative corrections are given for
several intervals of $W^\pm$ scattering angles. At high energies large
cancellations occur between negative virtual corrections and positive
corrections corresponding to real photon or $Z$-boson emission.
Consequently, although the correction originating from the $WWZ$
production is completely negligible at $\sqrt{s_{\g\g}}=0.3$~TeV, it
is of the same order of magnitude as hard photon correction at 2~TeV.
Although at $300\div 500$~GeV corrections are quite small ranging from
$-1.3\%$ to $-8\%$, depending on angular cuts, at TeV energies the
value of radiative corrections in the central region of $W^+W^-$
production become quite large, so that corrections in the region
$60^\circ < \theta < 120^\circ$ are $6\div 8$ times larger than the
corrections to the total cross section at $1\div 2$~TeV. They range
from $-24\%$ to $-45\%$. Thus if precision measurements are to be made
at TeV energy more careful theoretical analysis is needed in order to
reliably predict the value of the cross section in the central region
where the value of the cross section is the most sensitive to the $W$
anomalous couplings.

\section*{discussion}

Although the cross section of $WW$ production is much larger in $\g\g$
collisions, this fact itself is not to be considered as an obvious
advantage of PLC.  The reason is that although the anomalous
contribution to the amplitude of longitudinal $W_LW_L$ pair production
is enhanced by a factor of $s/M_W^2$ both in \GG\ for $J_z=0$ (for
$J_z=\pm 2$ in $\g\g$ collisions anomalous coupling contribution is
not enhanced at all) and $\epem$ collisions, the ${\cal SM}$ Born
amplitude of $W_LW_L$ production at PLC is suppressed as $M_W^2/s$, so
that the contribution of the interference term to the cross
section is decreasing as $1/s$ at PLC \cite{Trilinear}. On the
contrary, in $\epem$ collisions the anomalous contribution to the
cross section is enhanced, corresponding to non-decreasing cross
section of $W_LW_L$ production.  Recently the authors of Ref.
\cite{Trilinear} have demonstrated that enhanced coupling could still
be exploited in the \GG\ mode. Their clever idea is to reconstruct the
non diagonal elements of the $WW$ polarization density matrix by
analyzing the distributions of the decay products of the $W$'s,
thereby achieving the improvement over simple counting rate method of
more that an order of magnitude at $\rts=2$~TeV.  However, although
the benefits from \GG\ mode are really quite evident at $\rts=500$~GeV
and allowed region in the anomalous couplings parameter space shrinks
considerably when results from $e^+e^-\to W^+W^-$ are combined with
ones from $\g\g\to W^+W^-$, at energies above 1~TeV combining results
from $\epem$ and \GG\ modes does not considerably reduce the bounds
obtained from $\epem\to W^+W^-$ alone \cite{Trilinear}. This is
especially true for fits with one anomalous coupling. Qualitatively
these results can be understood considering the ratio $S/\sqrt{B}$ as
a measure of statistical significance of the anomalous coupling signal
$S$ with respect to the ${\cal SM}$ background $B$. Since the total
${\cal SM}$ cross section is decreasing as $1/s$ in $\epem$ collisions
and is constant in \GG\ collisions, while the enhanced anomalous cross
section behaves like a constant we get \be \frac{S(\epem\to
  W^+W^-)}{\sqrt{B(\epem\to W^+W^-)}} \propto \sqrt{s}, \ee while
$S(\g\g\to W^+W^-)/\sqrt{B(\g\g\to W^+W^-)} \propto 1$. If we take
into account that anomalous couplings affect mostly the cross section
in the central region, where the ${\cal SM}$ cross section behaves
like $\sigma(\g\g\to W^+W^-) \sim 8\pi \alpha^2/ p_T^2$, we get \be
\frac{S(\g\g\to W^+W^-)}{\sqrt{B(\g\g\to W^+W^-)}} \propto p_T, \ee
\ie\ the same improvement at higher energy as for $\epem$ collisions
but only for large values of $p_T$ cut $p_T\sim s$, with which the
cross section of $WW$ production in \GG\ collisions is not enhanced
any more with respect to production in $\epem$ collisions.  Moreover,
it is this region, where radiative corrections are as large as
$-24\div -45\%$.

To resume, at 500~GeV linear collider much stronger constraints on the
anomalous $W$ couplings could be obtained combining the bounds from
$\epem\to W^+W^-$ and $\g\g\to W^+W^-$. At such an energy radiative
corrections are well under control and photon-photon collider option
is essential for precision measurements.

At energies above 1~TeV, where enhancement factor $s/M_W^2$ for
anomalous amplidudes starts playing a crucial role, PLC is not so much
advantageous with respect to $e^+e^-$ collider at the same integrated
lumonosity. Radiative corrections are quite large and theoretical
consideration beyond one-loop approximation is needed in order to
safely predict ${\cal SM}$ cross section. Nevertheless, if in the
photon-photon collisions one can obtain much larger luminosity than in
$e^+e^-$ collisions \cite{Telnov97}, the PLC potential could still be as
high as at lower energies.

\section*{acknowledgments}

I am grateful to the organizers of the Workshop for warm hospitality
and to F.~Boudjema, I.~Ginzburg and V.~Telnov for fruitful discussions.
This work has been supported by the Alexander von Humboldt Foundation.


\begin{references}

\bibitem{Telnov97}
V.~Telnov, Proc. of the {\it International Conference on the
Structure and Interactions of the Photon, PHOTON '97}, 10--15 May
1997, Egmond aan Zee, The Netherlands, physics/9706035.

\bibitem{Jikia97}
G.~Jikia, ibid., hep-ph/9706508.

\bibitem{NLC}
NLC ZDR Design Group and NLC Physics Working Group (S. Kuhlman et al.), 
SLAC-R-0485, June 1996, hep-ex/9605011.

\bibitem{CDR} 
R.~Brinkmann {\it et al.}, DESY-79-048, July 1997, hep-ex/9707017.

\bibitem {Ano}  S.Y.Choi, F.Shrempp, Phys.Lett. {\bf B272} (1991)
149; E.Yehudai, Phys. Rev. {\bf D44} (1991) 3434; A.Miyamoto in Proc.
2-nd Workshop on JLC, KEK 91-10 (1991). 

\bibitem{BeBu}
G.~B\'elanger and F.~Boudjema, {\it Phys. Lett.} {\bf B288} (1992) 210.

\bibitem{BB}
M. Baillargeon and F. Boudjema {\it Phys. Lett.} {\bf B317} (1993) 371.

\bibitem{BBB}
M.~Baillargeon, G.~Belanger, and F.~Boudjema, Proc. of the {\it
``Two-Photon Physics from $DA\Phi NE$ to LEP200 and Beyond''}, 2-4 February
1994, Paris.

\bibitem{eboli}
F.T.~Brandt, O.J.P.~\'Eboli, E.M.~Gregores, M.B.~Magro, 
P.G.~Mercadante and S.F.~Novaes, {\it Phys. Rev.} {\bf D50} (1994)
5591.

\bibitem{gamma-gamma2} 
M.~Baillargeon, G.~B\'elanger, F.~Boudjema and G.~Jikia, 
Proceedings of the 3rd Workshop {\it $e^+e^-$ Collisions at
TeV Energies: the Physics Potential, Part D}, DESY, Hamburg, Germany,
August 30 -- September 1, 1995, p. 511.

\bibitem{DDS-WW}
A.~Denner, S.~Dittmaier, and R.~Schuster, 
{\it Nucl. Phys.} {\bf B452} (1995) 80;
Report BI-TP 96/03, WUE-ITP-96-001, hep-ph/9601355.

\bibitem{Jikia-WW}
G.~Jikia, Proc.  of the Workshop {Physics and
Experiments with Linear $e^+e^-$ Colliders}, Morioka, Japan, 1995;
{\it Nucl. Phys.} {\bf B494} (1997) 19.

\bibitem{Trilinear} M.~Baillargeon, G.~B\'elanger and F.~Boudjema,
preprint ENSLAPP-A-639/97, hep-ph/9701372.

\end{references}
\end{document}